\documentclass[%
 superscriptaddress,
 amsmath,amssymb,
 prx, nofootinbib,
 twocolumn
]{revtex4-2}
\usepackage{graphicx} 
\usepackage{comment}

\usepackage{dsfont}
\usepackage{xcolor}
\usepackage{bm} 
\usepackage{hyperref}
\usepackage{color,soul}

\usepackage[utf8]{inputenc}
\usepackage{amsmath}
\usepackage{amssymb}
\usepackage[a4paper, total={6.6in, 8.8in}]{geometry}
\usepackage{float}
\usepackage{braket}
\usepackage{adjustbox}
\usepackage{framed} 
\usepackage{url}
\usepackage[normalem]{ulem}
\usepackage{dcolumn}
\usepackage{bm}
\usepackage{physics}
\usepackage{tikz}

\definecolor{eggplant}{RGB}{180,33,147}

\makeatletter
\def\@hangfrom@section#1#2#3{\@hangfrom{#1#2}#3}
\def\@hangfroms@section#1#2{#1#2}
\makeatother

\begin{document}

\title{Confinement in the Transverse Field Ising model on the Heavy Hex lattice}

\author{Joseph Tindall}
\affiliation{Center for Computational Quantum Physics, Flatiron Institute, New York, New York 10010, USA}

\author{Dries Sels}
\affiliation{Center for Computational Quantum Physics, Flatiron Institute, New York, New York 10010, USA}
\affiliation{Center for Quantum Phenomena, Department of Physics, New York University, 726 Broadway, New York, NY, 10003, USA}

\date{\today}

\begin{abstract}
We study the emergence of confinement in the transverse field Ising model on a decorated hexagonal lattice. Using an infinite tensor network state optimised with belief propagation we show how a quench from a broken symmetry state leads to striking nonthermal behaviour underpinned by persistent oscillations and saturation of the entanglement entropy. We explain this phenomenon by constructing a minimal model based on the confinement of elementary excitations, which take the form of various flavors of hadronic quasiparticles due to the unique structure of the lattice. Our model is in excellent agreement with our numerical results. For quenches to larger values of the transverse field and/or from non-symmetry broken states, our numerical results displays the expected signatures of thermalisation: a linear growth of entanglement entropy in time, propagation of correlations and the saturation of observables to their thermal averages. These results provide a physical explanation for the unexpected simulability of a recent large scale quantum computation.
\end{abstract}

\maketitle

\textit{Introduction} - The mechanism by which an isolated many-body quantum system is able to relax and reach `equilibrium' is now relatively well understood \cite{Deutsch1991, Srednicki1994, Rigol2008, Hyungwon2014,dalessio2016, Vincenzo2015, Deutsch2018}. Following the eigenstate thermalization hypthesis (ETH), the relaxation of a local observable occurs because diagonal matrix elements in the energy eigenbasis are approximately constant over some energy shell of the Hilbert space, whereas off-diagonal matrix element are random. Most recently, by moving to a description in terms of eigenoperators instead of eigenstates the \textit{weak} form of the ETH (which makes statements on the time-averages of observables) was rigorously proven \cite{Buca2023}.
\par Naturally, an understanding of when a system thermalizes is fundamental to understanding when it does not, and stable, correlated phases of matter with non-trivial properties can emerge \cite{Iemini2018, Tindall2020superconductivity, Else2020, Bahri2015}. There are now many known pathways by which a quantum system can avoid its anticipated relaxation: `atypical' eigenstates which form scars in the eigenspectrum of the system \cite{Zhao2020, Serbyn2021, Turner2018, Zhang2023}, dynamical symmetries which generate a closed operator algebra in a subspace of the Banach space \cite{Buca2019, Tindall2020, Medenjak2020} and the existence of an extensive set of integrals of motion \cite{Konstantinidis2015} are but a handful.   
\par One of the simplest many-body models where athermal behaviour occurs on incredibly long timescales is the one-dimensional transverse field Ising (TFI) model in the presence of a small longitudinal field of strength $h_{\parallel}$ \cite{Birnkammer2022, Marton2017, Lake2010, Robinson2019, Shriya2020, Scopa2022, Mazza2019}. For $h_{\parallel} = 0$ the model is free-fermionic and, for small transverse field strength, the low lying excitations above the ground state are domain walls. The longitudinal field then acts as a confining potential on these quasiparticles, resulting in bound states of pairs of domain walls (often referred to a mesons, in analogy to how quarks get confined in quantum chromodynamics \cite{Tin2017}). Under a quench, these mesons are created and remain confined on extremely long timescales. At small $h_{\parallel}$ their decay is non-perturbative in $h_{\parallel}$. This confinement manifests itself through stable, persistent oscillations in the magnetisation and the entanglement entropy. Such behaviour has also been proposed in the two-dimensional transverse-field Ising model \cite{Ramos2020, James2019}, even in the absence of a longitudinal field. Here, it was shown that a weak perpendicular coupling between one-dimensional chains acts like a longitudinal field for a polarized initial state. Consequently, the resulting dynamics is analogous to the one-dimensional case. This simple coupled wire construction highlights how non-thermal behavior can arise after quantum quenches inside a symmetry broken phase. Such a situation can not arise in local one-dimensional models.

\begin{figure*}[t!]
    \includegraphics[width = \textwidth]{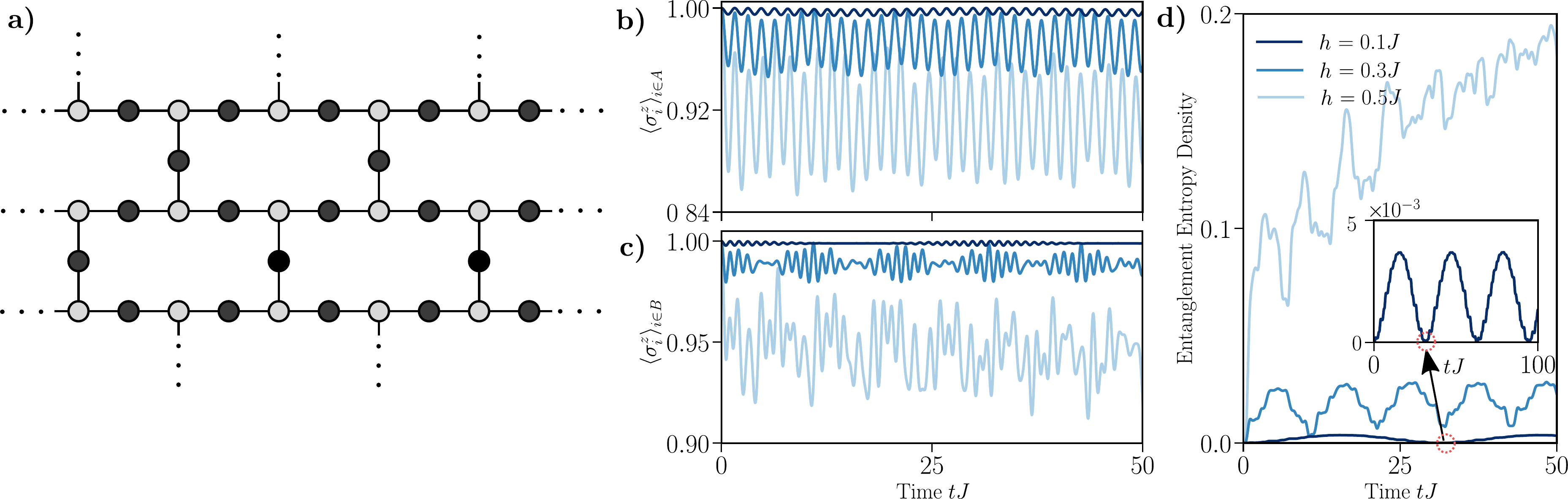}
    \caption{\textbf{a)} Finite region of the heavy hexagon lattice. The lattice is bi-partite and cosists of two sublattices $A$ (dark circles) and $B$ (light circles) which host sites of co-ordination number two and three respectively. \textbf{b-d)} Dynamics for a quench from the initial state  $\vert Z^{+} \rangle = \vert \uparrow \uparrow \uparrow ... \uparrow \rangle$ for a range of values of $h/J$ (from darkest to lightest, $h/J = 0.1,0.3,0.5$. \textbf{b)} Dynamics of the magnetisation on the $A$ sublattice. \textbf{c)} Dynamics of the magnetisation on the $B$ sublattice. \textbf{d)} Dynamics of $s$, the density of the entanglement entropy. The entanglement entropy for a given partition scales as $\mathcal{O}(sL)$ where $L$ is the size of the boundary between the partition and the rest of the lattice. The inset shows the dynamics of $s$ for $h/J = 0.1$ up to $tJ \leq 100$.}
    \label{Fig:Lattice}
\end{figure*}

\par In this paper we study confinement under a quench from a symmetry broken state in a fully two-dimensional system: the TFI model on a decorated hexagonal lattice. Here the coupling between spins is isotropic and the lattice has additional structure (two sublattices of different co-ordination numbers). The confining behaviour we observe is thus distinct, being underpinned by multiple species of hadronic quasiparticles of varying size and content. The lattice structure itself acts as the confining potential on these excitations, as increasing their size causes them to contain more domain walls. Performing numerical calculations on an infinite Tensor Network State (iTNS) optimised with belief propagation we perform accurate simulations which reveal the confined behaviour of the system in the thermodynamic limit. Our numerical results demonstrate excellent agreement with our minimal model of multiple zero-momentum hadronic quasiparticles.  For larger values of the transverse field and non-symmetry broken initial states, our numerical method yields the expected signatures of thermalisation in line with the predictions of the ETH.

\textit{Model and Numerical Results} - We consider the heavy hexagonal lattice: a regular hexagonal lattice where each edge is decorated with an additional lattice site. Unless otherwise stated all of our results are in the limit of an infinite number of lattice sites. A view of a finite region of the heavy hex lattice is shown in Fig. (\ref{Fig:Lattice}). The lattice is bi-partite with two sublattices $A$ and $B$ which contain sites of co-ordination number two and three respectively. We are interested in the transverse field Ising Hamiltonian (setting $\hbar = 1$ throughout)
\begin{equation}
    H = -J\sum_{\langle ij \rangle}\sigma^{z}_{i}\sigma^{z}_{j} + h\sum_{i}\sigma^{x}_{i},
    \label{Eq:TFIHamiltonian}
\end{equation}
with $\sigma^{\alpha}_{i}$ the relevant Pauli matrix acting on lattice site $i$. The first summation in Eq. (\ref{Eq:TFIHamiltonian}) runs over the pairs of nearest neighbours on the lattice and the second summation over all sites of the lattice. 
\par We focus on a quench under $H$ from the `symmetry-broken' initial state $\ket{ Z^{+} } = \ket{ \uparrow \uparrow \uparrow ... \uparrow }$,
one of the two degenerate ground states for $h = 0$. Our numerical method is to use an infinite tensor network state optimized with belief propagation (BP-iTNS). We trotterise the exponential $U(t) = \exp(- {\rm i}H t)$, and apply the resulting gates to the iTNS, performing truncations and taking expectation values using belief propagation \cite{TindallGauging2023, Tindall2023, TensorNetworkBeliefPropagation1, TensorNetworkBeliefPropagation2, TensorNetworkBeliefPropagation3} --- which operates under the assumption of tree-like correlations (rank-one environments) in the tensor network (see Supplemental Material, SM for full details of our numerical method \cite{SM}). The BP-iTNS method has already been shown to be extremely accurate \cite{Tindall2023, Chan2023, Liao2023} for modelling the discrete dynamics of a kicked heavy-hex system: a system which was originally simulated using the IBM Eagle quantum processor \cite{kim2023}. In our results we will provide numerical and analytical evidence for the accuracy of BP-iTNS in simulating the continuous-time dynamics generated by $U(t)$.

\begin{figure}[t!]
    \includegraphics[width = \columnwidth]{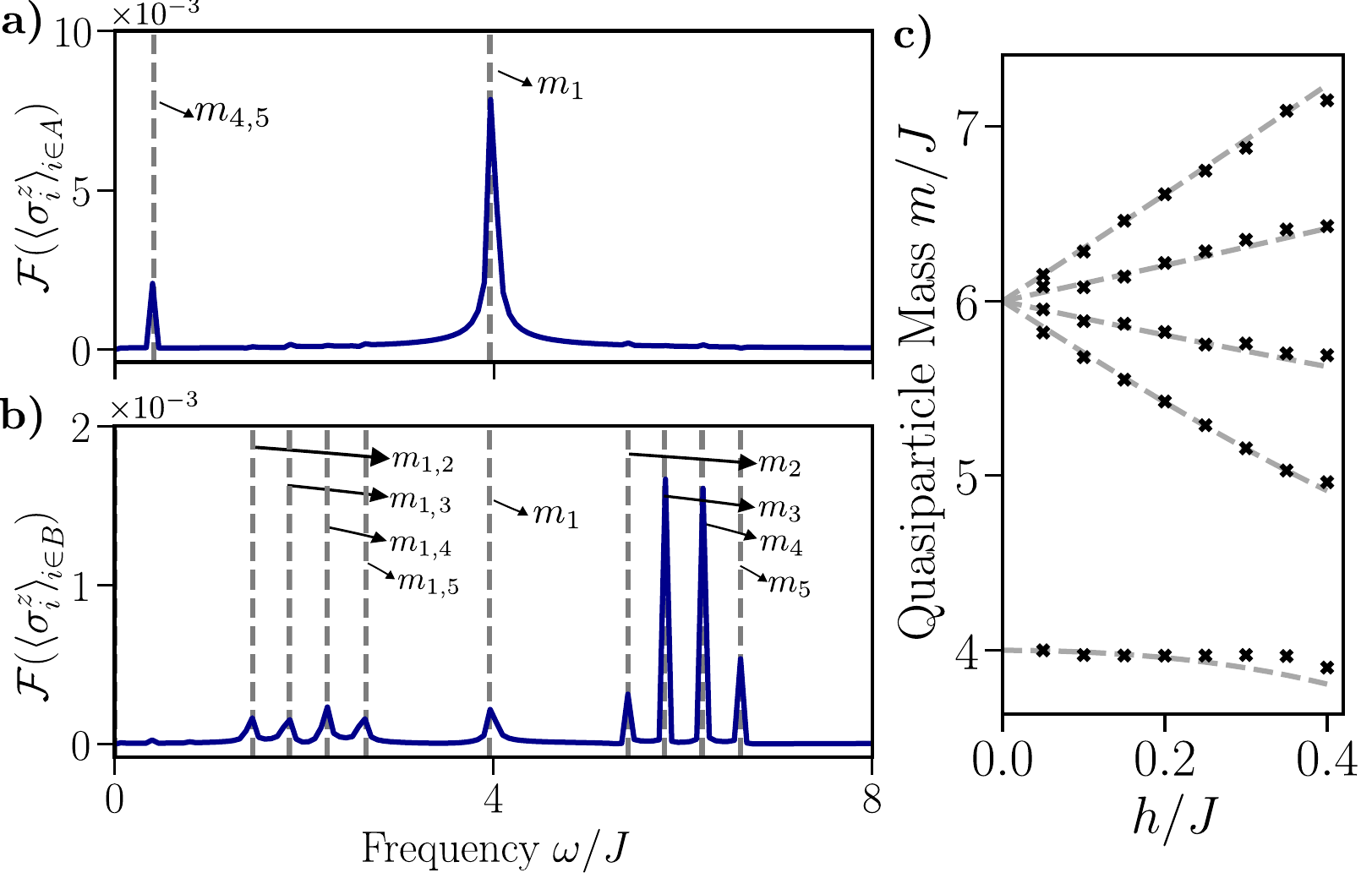}
    \caption{\textbf{a-b)} Fourier transform for $10 \leq tJ \leq 100$ of the magnetisation on the $A$ and $B$ sublattices of the infinite heavy hexagon lattice. The initial state is $\ket{ Z^{+}}  = \ket{ \uparrow \uparrow \hdots \uparrow }$ and is evolved under the Hamiltonian in Eq. (\ref{Eq:TFIHamiltonian}) with $h = 0.2J$. Grey dotted lines indicate the quasiparticle masses $m_{i}$ (and mass differences $m_{i,j} = \vert m_{i} - m_{j} \vert $) extracted from the eigenvalues of the confinement Hamiltonian in Eq. (\ref{Eq:ReducedHamiltonian}). \textbf{c)} Masses (dashed grey lines) extracted from the eigenvalues of the confinement model in Eq. (\ref{Eq:ReducedHamiltonian}) for a range of $h/J$. Data points represent masses extracted from the Fourier transform of the dynamics of the magnetisation calculated using iBP-TNS for $10 \leq tJ \leq 100$.} 
    \label{Fig:Excitations}
\end{figure}

\par In Figure \ref{Fig:Lattice} we show our results for a quench under $H$ for several values of $h$ up to $h \leq 0.5J$. For the values $h = 0.1J$ and $h = 0.3J$ we see a striking lack of thermalisation, with long-lived stable oscillations appearing in both the local magnetisation and the entanglement density $s$. The actual entanglement entropy for a given partition scales as $\mathcal{O}(sL)$ where $L$ is the size of the boundary between the partition and the rest of the lattice. Moreover, for $h = 0.1J$ we observe oscillations which return the system \textit{almost} exactly back to its original product state with zero entanglement. We wish to stress that this simulation is done entirely in the thermodynamic limit. For $h = 0.5J$ in Fig.~\ref{Fig:Lattice}, we still observe persistent oscillations in the magnetisation but it is decaying slowly and there is also a clear linear growth in the entanglement entropy with time: indicating that the system is slowly thermalising on the observation timescale. 
\par In the Supplementary Material (SM), we demonstrate how our iBP-TNS results in this small $h$ regime show strong agreement with time-dependent variational principle (TDVP) calculations on a Matrix Product State (MPS). This latter method does not make the assumption of tree-like correlations \cite{Jutho2011, Kloss2018} but only works on a finite-sized system and is more computationally expensive than iBP-TNS due to the mapping to one-dimension introducing long-range terms which causes an explosive growth in the bond dimension of the MPS. Nonetheless, for small $h$ it is possible to obtain accurate results with it and the observed agreement between the two methods serves as evidence for the accuracy of our iBP-TNS method and the validity of the assumption of tree-like correlations in the small $h$ regime. Moreover, BP has already been shown to be extremely accurate for modelling the discrete time-dynamics realised by a kicked Ising model on the same lattice \cite{Tindall2023}, performing favorably against complementary methods such as direct simulation on a quantum processor \cite{kim2023}, Clifford perturbation theory \cite{Chan2023} and matrix product operator simulation \cite{Zalatel2023}.
\par The behaviour observed in Fig. \ref{Fig:Lattice} is reminiscent of the confinement observed in one-dimensional transverse field Ising chains \cite{Birnkammer2022, Marton2017, Lake2010, Robinson2019, Shriya2020, Scopa2022, Mazza2019, Fangli2019, Ramos2020, James2019}. Low energy excitations in the form of mesonic quasiparticles (bound pairs of domain walls) are formed on short time-scales under a quench and, due to the presence of a confining potential for these mesons, the system remains, for long-times, trapped in the subspace associated with a small, non-interacting density of these quasiparticles. The confining potential can manifest itself either explicitly in the Hamiltonian via a small longitudinal field \cite{Marton2017} or implicitly via terms such as long-range interactions  \cite{Fangli2019} or a weak vertical coupling to other one-dimensional chains \cite{Ramos2020, James2019}.
\par Here, we observe such oscillations in the heavy-hexagonal lattice. Due to the additional structure in the lattice and the homogeneous nature of the spin-spin couplings, there are multiple low energy excitations which are relevant to the dynamics of the system and constitute different types of hadronic quasiparticles. These are excited by the quench and remain confined on long time-scales. We will we build up a minimal model of the dynamics for small $h$ based on this picture and show how it accounts, very accurately for the dynamics observed in Fig. \ref{Fig:Lattice}.

\textit{Confinement Model} - We consider the low energy excitations on top of the initial state $\vert Z^{+} \rangle$. When $h = 0$ the lowest energy excitation consists of flipping a single spin on sublattice $A$ at an energy cost of $4J$ as it creates two domain walls each with an energy penalty of $2J$. We notate the basis spanned by the states containing these single excitations with $\mathcal{A} = \{\vert i_{A} \rangle \}$ where $i_{A}$ indexes the position of the flipped spin on the $A$ sublattice. The excitations with the second lowest energy cost $6J$ and involve flipping a single spin on the $B$ sublattice and some subset $I_{A}$ of the set of its neighbors $n(i_{B})$, creating $3$ domain walls. We notate these with the basis states $\{ \vert i_{B}, I_{A} \subseteq n(i_{B}) \rangle \}$. We use $\mathcal{B}(\vert I_{A} \vert)= \{ \vert i_{B}, I_{A} \subseteq n(i_{B}) \rangle_{\vert I_{A} \vert} \}$ to indicate the set of basis states with a fixed size for the subset of neighbors $ \vert I_{A} \vert$. Clearly $0 \leq \vert I_{A} \vert \leq 3$, giving us, in total, five different sets of basis states which make up the excitations with a gap less than or equal to $6J$ above the ground state in the system for $h = 0$.

\par Let us now define the zero-momentum state in a given set of basis states $\mathcal{S}$
\begin{equation}
    \ket{ \mathcal{S}_{ \vert \Vec{k} \vert = 0} } = \frac{1}{\sqrt{\vert \mathcal{S} \vert}}\sum_{s \in \mathcal{S}}\ket{ s }. 
\end{equation} 
The projector into the zero-momentum basis consisting of the five different low energy excitations is then
\begin{equation}
  \mathcal{P}_{\vert \Vec{k} \vert = 0} = \ket{ \mathcal{A}_{ \vert \Vec{k} \vert = 0} } \bra{ \mathcal{A}_{ \vert \Vec{k} \vert = 0} }  + \sum_{i = 0}^{3} \ket{ \mathcal{B}(i)_{ \vert \Vec{k} \vert = 0} } \bra{ \mathcal{B}(i)_{ \vert \Vec{k} \vert = 0} }.
\end{equation}

\begin{figure}[t!]
    \includegraphics[width = \columnwidth]{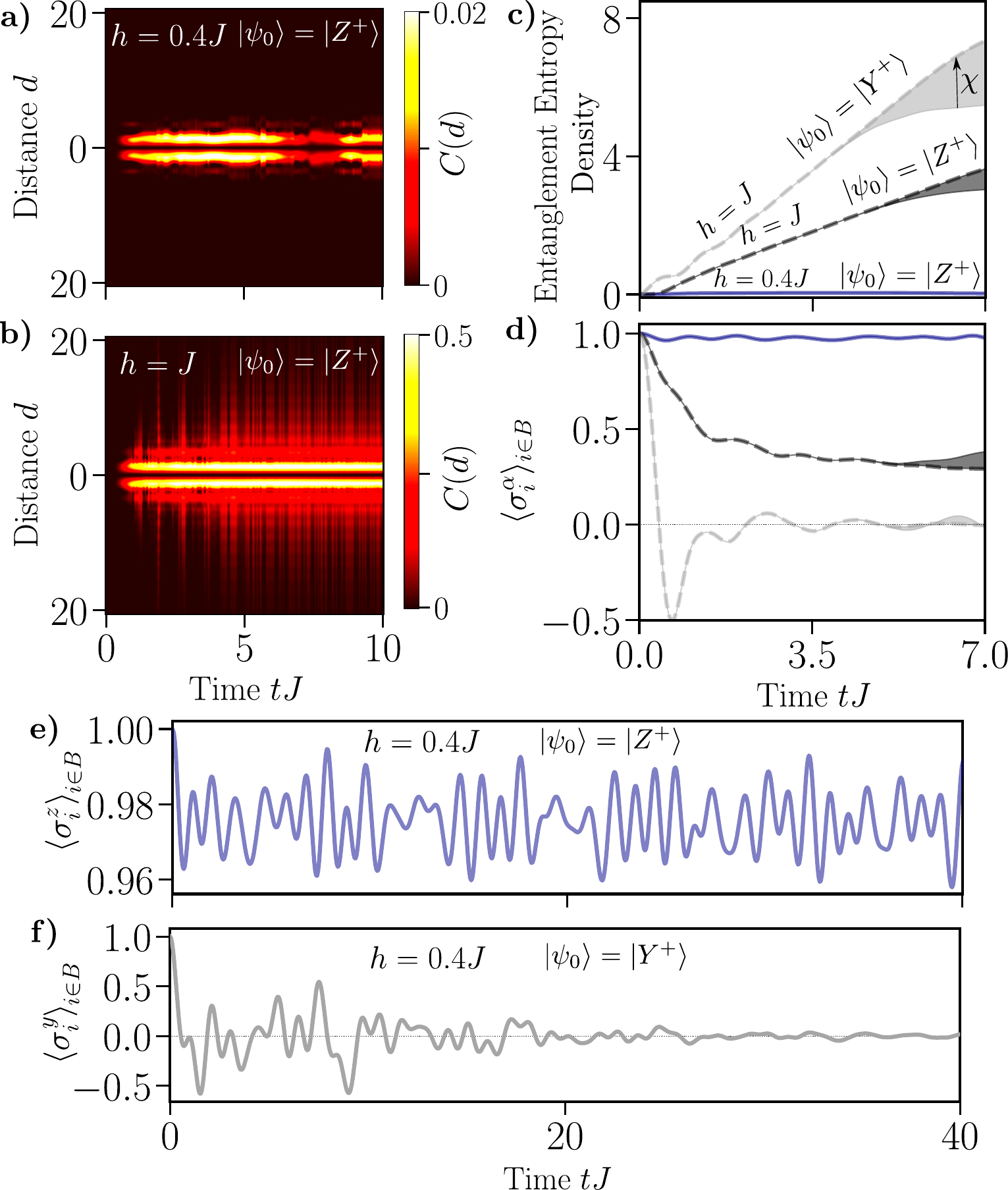}
    \caption{\textbf{a-b)} Dynamics of the two-point correlator $C(d) = \langle \sigma^{z}_{i}\sigma^{z}_{i+d} \rangle -\langle \sigma^{z}_{i} \rangle \langle \sigma^{z}_{i+d} \rangle$ as a function of distance $d$ and time $tJ$ for a quench from the initial state $\ket{\psi_{0} } = \ket{ Z^{+} }$ for $h = 0.4J$ (top) and $h = J$ (bottom). We choose the site $i$ to be in the $A$ sublattice. The correlator is calculated under the BP-approximation (see SM for details). \textbf{c-d)} Dynamics of the entanglement entropy and of the on-site magnetisation. The magnetisation is measured along the spin-axis $\alpha$ the initial state is polarised. Solid blue line corresponds to $\ket{\psi_{0}} = \ket{ Z^{+} }$ and $h = 0.4J$, dark-grey dashed line is $\ket{\psi_{0}} = \ket{ Z^{+} }$ and $h = J$ whilst light-grey dashed line is $\ket{\psi_{0}} = \ket{ Y^{+} }$ and $h = J$. The shaded area represents the range of results for bond dimensions $50 \leq \chi \leq 300$ of our BP-optimised iTNS (with the dashed line representing the largest bond dimension used). \textbf{e-d)} Dynamics of the on-site magnetisation for the initial states $\ket{ Y^{+} }$ and $\ket{ Z^{+} }$ for $h = 0.4J$. The magnetisation is measured along the spin-axis the initial state is polarised.} 
    \label{Fig:Thermalisation}
\end{figure}

Projecting the Hamiltonian in Eq. (\ref{Eq:TFIHamiltonian}) into this `confined', zero-momentum basis gives us (see SM for derivation \cite{SM}) the following effective Hamiltonian
\begin{equation}
    \mathcal{P}_{\vert \Vec{k} \vert = 0} H \mathcal{P}_{\vert \Vec{k} \vert = 0} = \begin{pmatrix}
        4J & 0 & \sqrt{2} h & 0 & 0 \\
        0 & 6J & \sqrt{3}h & 0 & 0 \\
        \sqrt{2}h & \sqrt{3}h & 6J &  0 & 2h \\
        0 & 0 & 0 & 6J & \sqrt{3}h \\
        0 & 0 & 2h & \sqrt{3}h & 6J
    \end{pmatrix},
    \label{Eq:ReducedHamiltonian}
\end{equation}

for \textit{any} sized heavy hexagon lattice with periodic boundary conditions. The ordered (smallest to largest) eigenvalues $m_{1}, m_{2}, ..., m_{5}$ of this Hamiltonian define the masses (in units of $J$, as we have set $\hbar = 1$) of our quasiparticle excitations.

\par In Fig. \ref{Fig:Excitations} we plot the Fourier transform of the dynamics of the magnetisation of the infinite system for $h = 0.2J$, which were calculated using iBP-TNS. We compare the frequencies present against the masses calculated from our model. We observe excellent agreement between the two, indicating our picture of confined, low energy quasiparticles is correct and further validating the assumption of tree-like correlations in the lattice (as our confinement model does not make this assumption). In Fig. \ref{Fig:Excitations}c we compare the masses against the amplitudes from the Fourier transform of the iBP-TNS dynamics for a range of values of $h$, observing good agreement. Discrepancies between the two become more noticeable as $h$ increases, indicating our minimal model of confinement has to be extended with higher energy excitations. This is consistent with the visible slow thermalisation observed in Fig. \ref{Fig:Lattice} for $h = 0.5J$.

\textit{Thermalisation} - In Fig. \ref{Fig:Thermalisation} we compare the behaviour of the system under confinement to when it is thermalising. To supplement our analysis and consider the spread of information in the system we calculate, under the BP approximation, the two-point correlator $C(d) = \langle \sigma^{z}_{i}\sigma^{z}_{i+d} \rangle -\langle \sigma^{z}_{i} \rangle \langle \sigma^{z}_{i+d} \rangle$ as a function of time and distance $d$ (see SM \cite{SM} for calculation details). Here we define distance $d$ as the smallest length path (on the lattice) between two sites and, under the BP approximation, this correlator depends only on $d$ and the sublattice which $i$ belongs to. For the initial state $\vert Z^{+} \rangle$ and $h = 0.4J$ we observe a clear signature of confinement with the correlations remaining completely localised and relevant only up to very short distances. This is corroborated by persistent oscillations in the magnetisation and an extremely slow growth in the entanglement entropy. For $h = 1.0J$ we instead observe the anticipated signatures of thermalisation: the spreading of information (correlations) in time, a linear growth in entanglement entropy and the saturation of the on-site expectation value $\langle \sigma^{z}_{i \in B} \rangle$. 
\par We also consider a different initial state which is not symmetry-broken $\vert Y^{+} \rangle$, where all spins are polarised along the $y$ spin axis. For $h = 0.4J$ and $h = 1.0J$ the magnetisation $\langle \sigma^{y}_{i \in B} \rangle$ decays and settles around zero. This is consistent with the prediction from the ETH/ Gibbs ensemble: $\langle Y^{+} \vert H \vert Y^{+} \rangle = 0 \ \forall h$ and so local observables for the initial state should relax, under a quench, to those consistent with an infinite temperature thermal state $\rho \propto \mathbb{I}$ as it also has zero energy (${\rm Tr}(H) = 0$ due to the eigenspectrum being symmetric about $0$). We highlight the stark contrast between the dynamics of the initial states $\vert Z^{+} \rangle$ and $\vert Y^{+} \rangle$ for $h = 0.4J$ in Fig. \ref{Fig:Thermalisation}e and Fig. \ref{Fig:Thermalisation}f, where the latter clearly relaxes whilst the former undergoes persistent oscillations with non-decaying amplitudes.
\par \textit{Truncating the iTNS} - In our simulations in the confinement regime ($h \leq 0.5J$ and $\ket{\psi_{0}} = \ket{Z^{+}}$ we enforce a \textit{cutoff} of $\epsilon \leq 10^{-10}$ (i.e. we discard the smallest singular values such that the sum of their squares is $\leq \epsilon$) when performing truncations on our iTNS \cite{itensor-r0.3}. With this we are able to keep up with the growth in bond-dimension of the iTNS throughout the evolution. Outside of the confinement regime, the faster growth of entanglement entropy compared to the confinement regime means we have to actively truncate our iTNS up to a maximum bond dimension $\chi$. We show the effect from using a range of bond dimensions in Fig. \ref{Fig:Thermalisation}. As we increase the bond dimension our results converge towards the behaviour predicted by the ETH and in the SM we show more detailed plots of the effect of increasing the bond dimension in this regime \cite{SM}.

\textit{Conclusion} - We have identified the emergence of confinement in the transverse field Ising model on the heavy-hexagon lattice. The more complex structure of the lattice and the isotropic nature of the spin-spin couplings means that, in comparison to previous work on the chain and square lattices, there are multiple types of quasi-particles which need to be accounted for to accurately identify the low energy degrees of freedom. We have presented a minimal model for the low energy physics of the system based on these quasi-particles and shown how it provides strong agreement with our state-of-the-art iBP-TNS calculations. 
\par Looking forward, we anticipate our methodology here for deriving the relevant low energy degrees of freedom can be used to analyse the role of confinement on a range of physical lattice structures. Whilst confinement has been studied extensively on one-dimensional systems, little is known about its emergence on more complicated lattice structures. Our results here represent an initial foray into this space.
\par Finally, exceptionally slow thermalising behaviour and the growth of entanglement was recently observed in discrete-time realisations of the transverse field Ising dynamics on a heavy-hexagon quantum processor \cite{Tindall2023, kim2023}. Our results here indicate the physical origin of this behaviour is confinement and also explain why the BP-iTNS method, and a plethora of other methods \cite{Zalatel2023, begušić2023, Chan2023, kechedzhi2023, Liao2023} were capable of simulating the physics of the heavy-hexagon quantum processor with such ease. The emergence of confinement however, is also the reason why the physical observables measured in the experiment in Ref.~\cite{kim2023} took on non-trivial values away from the anticipated thermal predictions. Without confinement, local expectation values would have quickly been consistent with their thermal values.

\begin{acknowledgements}
    J.T. and D.S. are grateful for ongoing support through the Flatiron Institute, a division of the Simons Foundation. D.S. was supported by AFOSR: Grant FA9550-21-1-0236. We would like to acknowledge Matt Fishman, Miles Stoudenmire and Berislav Buca for insightful discussions. The code used to produce the numerical results in this paper was written using the \textbf{ITensorNetworks.jl} package \cite{ITensorNetworks} --- a general purpose and publicly available Julia \cite{bezanson2017julia} package for manipulating tensor network states of arbitrary geometry. It is built on top of \textbf{ITensors.jl} \cite{itensor-r0.3}, which provides the basic tensor operations. 
\end{acknowledgements}

\clearpage
\newpage
\onecolumngrid

\renewcommand{\theequation}{S\arabic{equation}}
\renewcommand{\figurename}{Supplementary Figure}
\setcounter{equation}{0}
\setcounter{figure}{0}     

\section*{Supplementary Material for: Confinement in the Transverse Field Ising model on the Heavy Hex lattice}
\section*{Numerical Details: iBP-TNS Method}

\begin{figure*}[t!]
    \includegraphics[width = \textwidth]{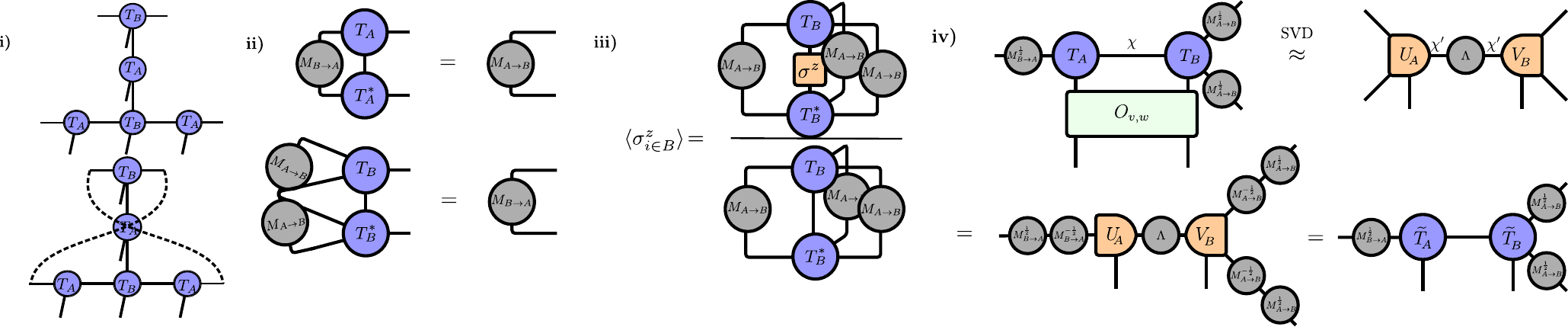}
    \caption{\textbf{i)} Top) Tensor Network State (TNS) over a unit cell of the infinite heaxy hexagonal lattice. The symmetry of the lattice is characterised by the two distinct tensors $T_{A}$ and $T_{B}$ which have two and three virtual bond indices respectively. Bottom) TNS optimised during our simulations. The dotted lines represent edges added to the unit cell create the necessary periodic boundary conditions for the simulation to map to the results of the thermodynamic limit. \textbf{ii)} The two unique belief propagation equations for updating the two possible message tensors (which represent rank-$1$ approximations of the environments for the TNS). \textbf{iii)} Calculation of an on-site expectation value on the $B$ sublattice using the message tensors. A similar equation can be defined for sublattice $A$. \textbf{iv)} `Simple update' procedure for applying a two-site gate on an edge of the TNS for the infinite heavy-hex lattice using the square roots of the message tensors found via belief propagation. Further details can be found in Ref. \cite{TindallGauging2023}. This simple update procedure with message tensors (which works on a TNS in an arbitrary gauge) is analagous to the simple update procedure commonly applied in literature \cite{Jiang2008, Vlaar2021, Jahromi2020, Jahromi2021} to states in the Vidal gauge \cite{Shi2006, TindallGauging2023}.}
    \label{Fig:iTNSHeavyHex}
\end{figure*}
Our Tensor Network State (TNS) is the unit cell of the infinite heavy hex lattice (see Supplementary Fig. \ref{Fig:iTNSHeavyHex}i) with periodic boundary conditions \cite{TindallGauging2023}. The Hamiltonian from the main text is then defined over this periodic unit cell

\begin{equation}
    H = -J\sum_{\langle ij \rangle}\sigma^{z}_{i}\sigma^{z}_{j} + h\sum_{i}\sigma^{x}_{i},
    \label{Eq:TFIHamiltonianAppendix}
\end{equation}
and its exponential is trotterised
\begin{equation}
    U( t) = \exp(- {\rm i} H t ) = \left(\left(\prod_{\langle ij \rangle} \exp \left({\rm i} J \frac{\delta t}{2} \sigma^{z}_{i}\sigma^{z}_{j} \right) \right)\left( \prod_{i} \exp(-{\rm i} \delta t \sigma^{x}_{i} )  \right) \left(\prod_{\langle ij \rangle} \exp \left({\rm i} J \frac{\delta t}{2} \sigma^{z}_{i}\sigma^{z}_{j} \right) \right)\right)^{n}
\end{equation}
such that $t = n \delta t$ and $\delta t J \ll 1$ (we set $\delta t J = 0.05$ throughout our calculations). The gates in $U(t)$ are applied in sequence to the initial TNS which is always a product state with bond-dimension $1$. Single-site gates are applied directly and two-site gates are applied via the simple update procedure: using belief propagation on the TNS to identify the optimal rank-1 environments (see Supplementary Fig. \ref{Fig:iTNSHeavyHex}ii for the BP equations in this context) and applying the gate conditioned on these environments (see Supplementary Fig.~\ref{Fig:iTNSHeavyHex}iv and Ref. \cite{TindallGauging2023} for more details). Single-site expectation values are calculated using the belief propagation message tensors (see Supplementary Fig. \ref{Fig:iTNSHeavyHex}iii). Meanwhile, the `entanglement entropy density' is calculated as $\sum_{\lambda}\lambda^{2}\log_{2}(\lambda^{2})$ where $\lambda$ are the diagonal entries of the bond tensor (which are identical on each edge of the lattice) found by transforming the TNS to the `Vidal' gauge. This transformation can be done with the message tensors found from belief propagation \cite{TindallGauging2023}. Our method with this small unit cell immediately recovers the results from optimising the TNS of the infinite heavy-hex lattice with BP, hence why we refer to it as `BP-iTNS'.
\par
\begin{figure*}[t!]
    \includegraphics[width = \textwidth]{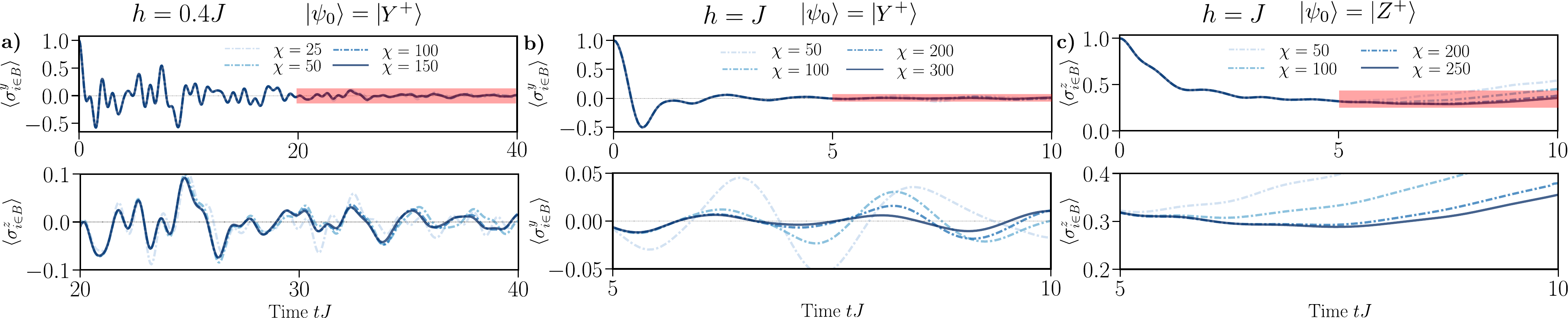}
    \caption{\textbf{a-c)} Convergence with bond dimension for results from Fig. 3 of the main text using the iBP-TNS method. \textbf{a)} Initial state is $\ket{\psi_{0}} = \ket{Y^{+}}$ and $h = 0.4J$. \textbf{b)} Initial state is $\ket{\psi_{0}} = \ket{Y^{+}}$ and $h = J$. The $x$-axis has been extended further than that used in the main text. \textbf{c)} Initial state is $\ket{\psi_{0}} = \ket{Y^{+}}$ and $h = J$. The $x$-axis has been extended further than that used in the main text. In all plots, solid blue lines reflect the value of $\chi$ used in the main text. Top plots have the $y$-scale identical to that used in the main text. Bottom plots represent zoomed in version of the top plots with the red boxes of the top plots reflecting the region zoomed in.} 
    \label{Fig:Fig3BondDimBehaviour}
\end{figure*}

 \textit{Truncation} - For the results starting in the initial state $\ket{ \psi_{0} } = \ket{Z^{+}}$ with $h \leq 0.5J$ in the main text we, after each SVD, discard away the smallest $n$ singular values such that the sum of their squares is $\epsilon \leq 10^{-12}$. With this truncation we are easily able to keep up with the growth in bond dimension of the TNS for the given timescales. We observe no discernible difference in our results when truncating with $\epsilon \leq 10^{-12}$ versus $\epsilon \leq 10^{-10}$. This indicates that our results are converged under the BP approximation in this regime. For different initial states and/or $h > 0.5J$ (which constitute our results in Fig. 3 of the main text) we enforce a maximum bond dimension in our truncations due to the faster growth in entanglement entropy in the thermalising regime. Nonetheless we find that our results are still converged in bond-dimension in comparison to the scale of the plots: Figure \ref{Fig:Fig3BondDimBehaviour} shows this explicitly, comparing the traces of expectation values for several different bond dimensions and showing how the differences are not discernible on the original scale of the plot --- only when zoomed in where we can see they are clearly converging.

\par
\begin{figure*}[t!]
    \includegraphics[width = \textwidth]{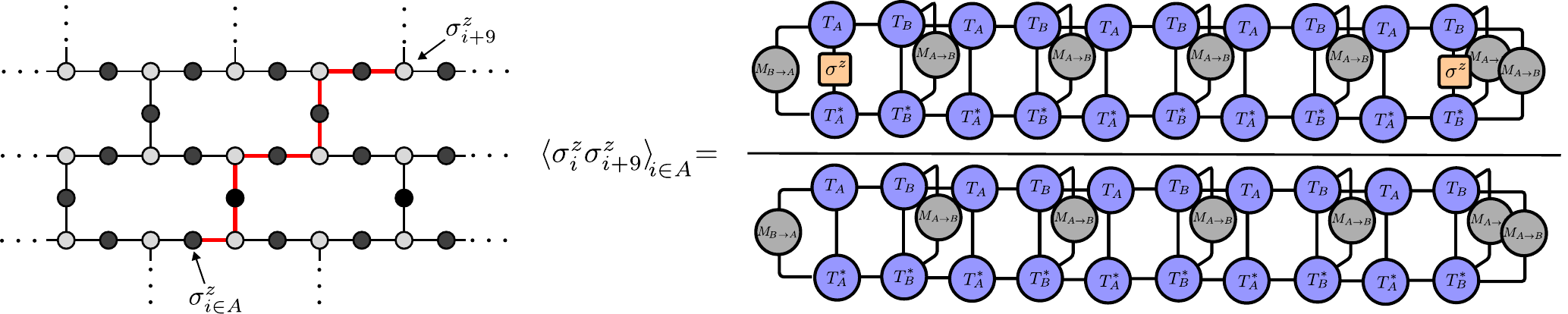}
    \caption{Method for calculating a two-point correlator $\langle \sigma^{\alpha}_{i} \sigma^{\beta}_{i + d} \rangle$ on the Heavy-Hex lattice using the iBP-TNS ansatz. The path between sites $i$ and $j = i + d$ on the infinite heavy-hex lattice is identified and a finite 1D TNS is constructed along that path using the on-site tensors $T_{A}$ and $T_{B}$ from the iBP-TNS ansatz. Message tensors found from belief propagation are applied as environments along the bonds of the lattice broken by that path. The result is the efficient calculation of any two-point correlator on the lattice. Here we show the explicit example of $\langle \sigma^{z}_{i} \sigma^{z}_{i + d} \rangle$ with $i \in A$.}
    \label{Fig:iTNSHeavyHexCorrelator}
\end{figure*}

 \textit{Two-Site Correlator} - Supplementary Figure \ref{Fig:iTNSHeavyHexCorrelator} shows how we compute the correlation $\langle \sigma^{z}_{i}\sigma^{z}_{i+d} \rangle$ from the iBP-TNS, where $i + d$ indicates a site at a distance (in terms of the path length through the lattice) $d$ from site $i$. Under the BP approximation for the environments of the iTNS, the correlator $\langle \sigma^{z}_{i}\sigma^{z}_{i+d} \rangle$ will depend only on the path length $d$ and which sublattice the site $i$ is in. Our method to calculate it is thus to consider the path between sites $i$ and $i + d$ on the infinite heavy-hex lattice and construct the TNS along that path using the tensors found in our optimisation of the periodic unit cell for the lattice. We apply the message tensors found from belief propagation as environments along the bonds of the lattice broken by that path. Using this method we can efficiently calculate, under the BP approximation, any two-point correlator on the lattice.

\section*{Numerical Details: Comparison to TDVP with an MPS}
As a further benchmark for our iBP-TNS method we also consider finite-realisations of the periodic heavy hex lattice (see Supplementary Fig.~\ref{Fig:TDVPvsBP}a), encoding the wavefunction as an MPS and the Hamiltonian as a long-range MPO. The system is then evolved using the TDVP (time-dependent variational procedure) and expectation values on a site in the centre of the lattice are calculated using the standard methods for an MPS. We order the sites of the MPS using a `snaking' through the heavy hex lattice which minimises the number of long-range terms \cite{Tindall2023}. We emphasize that without global subspace expansion \cite{Mingru2020} (which we do not use here) the TDVP method is extremely sensitive to the ordering of the lattice sites and this `snaking' is one of the few site orderings where we were able to obtain meaningful results.

\par For small $h/J$ and our `snaking' method we are able to keep up with the entanglement generated by the long-range MPO and thus achieve convergence: creating a benchmark for our results from iBP-TNS. We compare the two methods in Supplementary Fig. \ref{Fig:TDVPvsBP}. Strong agreement is seen between the two methods when we use a sufficiently large system size for the MPS calculation. We emphasize that convergence happens fairly rapidly in systems size. This suggests that both the BP-approximation is working extremely well for small values of $h/J$ and that finite-size effects from the lattice are minimal, presumably due to the confiment creating a low correlation length. For larger values of $h$ the entanglement grows too fast in the MPS calculations and the results are unable to serve as a meaningful benchmark for our iBP-TNS calculations (in the main text we instead relying on observing behaviour consistent with thermalization and the predictions of the ETH for a benchmark).

\begin{figure*}[t!]
    \includegraphics[width = \textwidth]{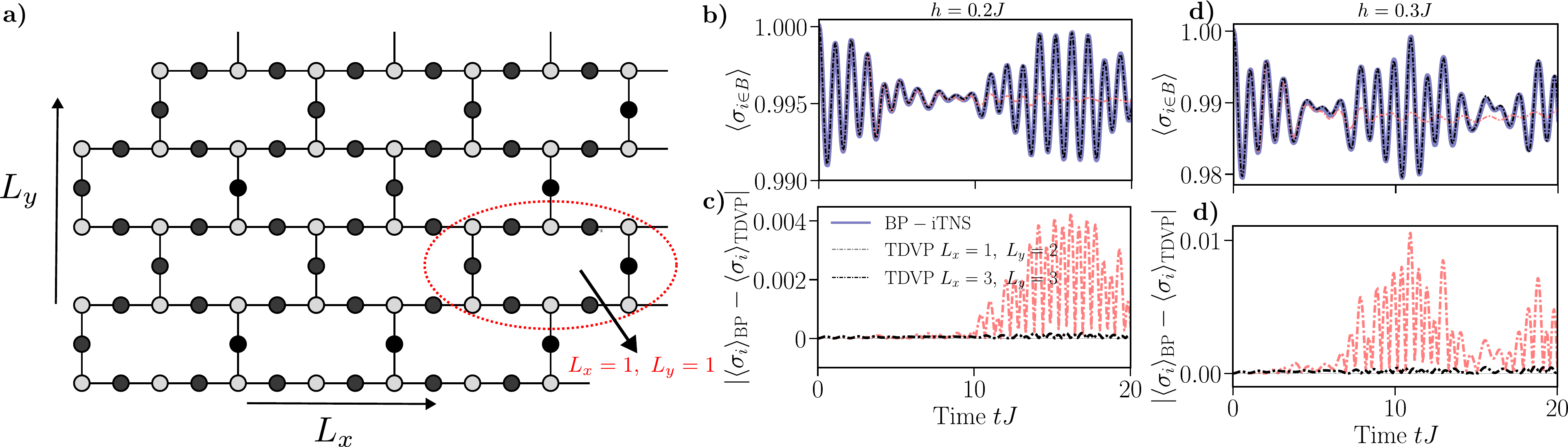}
    \caption{\textbf{a)} Heaxy-hexagon lattice formed of $L_{x}$ columns of heavy-hexagons and $L_{y}$ rows. The lattice is bi-partite and cosists of two sublattices $A$ (dark circles) and $B$ (light circles) which host sites with co-ordination numbers $2$ and $3$ respectively. The ringed region corresponds to $1$ heavy-hexagon and thus $L_{x} = 1$ and $L_{y} = 1$. \textbf{b-e)} Results for a quench from the initial state $\vert Z^{+} \rangle$ for $h = 0.2J$ and $h = 0.3J$. Solid blue line represents results (which are converged in bond dimension) using the iBP-TNS numerical method which acts directly in the thermodynamic limit. Dashed lines represents finite-size lattice results (which are also converged in bond dimension) using a Matrix Product State (MPS) ansatz and evolving with the Time Dependent Variational Principle (TDVP), where the Hamiltonian has been encoded as a long range MPO. We order the sites of the MPS using a `snaking' through the heavy hex lattice which minimises the number of long-range terms \cite{Tindall2023}. The magnetisation is measured on a site (in the B sublattice) in the middle of the lattice. \textbf{b-c)} Magnetisation on sublattice B and absolute difference in this magnetisation between the BP and TDVP results for $h = 0.2J$. \textbf{c-d)} Magnetisation on sublattice B and absolute difference in this magnetisation between the BP and TDVP results for $h = 0.3J$.  }
    \label{Fig:TDVPvsBP}
\end{figure*}

\section*{Deriving the confinement Hamiltonian}
Here we derive the form of the effective Hamiltonian $H_{\rm eff} = \mathcal{P}_{\vert \Vec{k} \vert = 0} H \mathcal{P}_{\vert \Vec{k} \vert = 0}$ given in the main text. This is the projection of the Transverse Field Ising Hamiltonian $H$ on \textit{any} heavy-hex lattice with periodic boundary conditions using the projector
\begin{equation}
  \mathcal{P}_{\vert \Vec{k} \vert = 0} = \vert \mathcal{A}_{ \vert \Vec{k} \vert = 0} \rangle \langle \mathcal{A}_{ \vert \Vec{k} \vert = 0} \vert  + \sum_{i = 0}^{3} \vert \mathcal{B}(i)_{ \vert \Vec{k} \vert = 0} \rangle \langle \mathcal{B}(i)_{ \vert \Vec{k} \vert = 0} \vert.
  \label{Eq:Projector}
\end{equation}
We have defined the normalized basis vectors
\begin{align}
    &\vert \mathcal{A}_{ \vert \Vec{k} \vert = 0} \rangle = \frac{1}{\sqrt{n_{A}}}\sum_{i_{A} \in A}\vert i_{A} \rangle, \notag \\
    &\vert \mathcal{B} \left(0 \right)_{ \vert \Vec{k} \vert = 0} \rangle = \frac{1}{\sqrt{n_{B}}}\sum_{i_{B} \in B}\vert i_{B} \rangle \notag \\
    &\vert \mathcal{B}\left( 1 \right)_{ \vert \Vec{k} \vert = 0} \rangle = \frac{1}{\sqrt{3n_{B}}}\sum_{i_{B} \in B} \ \sum_{j_{A} \in n (i_{B})} \vert i_{B}, j_{A} \rangle \notag \\
    &\vert \mathcal{B}\left( 2 \right)_{ \vert \Vec{k} \vert = 0} \rangle = \frac{1}{\sqrt{3n_{B}}}\sum_{i_{B} \in B} \ \sum_{j_{A}, k_{A} \in P(n (i_{B}))} \vert i_{B}, j_{A}, k_{A} \rangle, \notag \\
    &\vert \mathcal{B}\left( 3 \right)_{ \vert \Vec{k} \vert = 0} \rangle = \frac{1}{\sqrt{n_{B}}}\sum_{i_{B} \in B} \vert i_{B}, n(i_{B}) \rangle,
    \label{Eq:BasisStatesAppendix}
\end{align}
using $n_{A}$ to denote the number of sites in sublattice $A$ and $n_{B}$ the number of sites in sublattice $B$. The roman indices on the right hand side reference the lattice sites and when they appear in a given ket they refer to which spins are flipped with respect to the reference state $\vert Z^{+} \rangle = \vert \uparrow \uparrow \uparrow ... \uparrow \rangle$, which is the ground state for $h = 0$. The subscript of these indices refers to which sublattice the spins are on, $n(i_{B})$ denotes the set of $3$ neighbouring sites of site $i_{B}$. We have also introduced $P(n(i_{B}))$ to denote the set of all unordered pairs of the elements of $n(i_{B})$ and we use $j_{A}, k_{A} \in P(n(i_{B}))$ to index the two sites which form a given pair. Here we clearly have $|P(n(i_{B}))| = 3$. Finally, the argument for $\mathcal{B}$ on the LHS reflects the number of neighbours of the flipped spin $i_{B}$ that have also been flipped in the basis vectors that are summed over.
\par The state $\vert \mathcal{A}_{ \vert \Vec{k} \vert = 0} \rangle$ is the first zero-momentum excited state above the ground state with excitation energy $2J \times 2 = 4J$ due to each flipped spin on the $A$ sublattice having two neighbours with which they are then anti-aligned: thus creating two domains walls of energy $2J$. Meanwhile the remaining basis states are zero-momentum and have excitation energy $2J \times 3 = 6J$, due to every basis vector containing $3$ domain walls. We therefore have
\begin{align}
    \mathcal{P}_{\vert \Vec{k} \vert = 0} \left(-J\sum_{\langle ij \rangle}\sigma^{z}_{i}\sigma^{z}_{j} \right) \mathcal{P}_{\vert \Vec{k} \vert = 0} = {\rm Diag}(4J, 6J, 6J, 6J, 6J)
    \label{Eq:ReducedZZHamiltonian}
\end{align}
where, without loss of generality, we have set the energy of the state $\vert Z^{+} \rangle$ to $0$ and ordered the basis states as they appear (top to bottom) in Eq. (\ref{Eq:BasisStatesAppendix}).
\par Now we wish to determine the projection 
\begin{equation}
    \mathcal{P}_{\vert \Vec{k} \vert = 0} \left(h\sum_{i}\sigma^{x}_{i} \right) \mathcal{P}_{\vert \Vec{k} \vert = 0} = \mathcal{P}_{\vert \Vec{k} \vert = 0} \left(h\sum_{i}\left(\sigma^{+}_{i} + \sigma^{-}_{i} \right) \right) \mathcal{P}_{\vert \Vec{k} \vert = 0}
\end{equation}
The non-zero terms are clearly only those which couple basis states with have $\pm 1$ flipped spins with respect to each other. Consider for example the non-zero term 
\begin{equation}
    \langle \mathcal{B}(1)_{ \vert \Vec{k} \vert = 0} \vert h\sum_{i}\sigma^{x}_{i} \vert \mathcal{A}_{ \vert \Vec{k} \vert = 0}  \rangle = \langle \mathcal{B}(1)_{ \vert \Vec{k} \vert = 0} \vert h\sum_{i}\sigma^{+}_{i} \vert \mathcal{A}_{ \vert \Vec{k} \vert = 0}  \rangle.
\end{equation}
We have
\begin{equation}
    \left(\sum_{j}\sigma^{+}_{j} \right) \vert \mathcal{A}_{ \vert \Vec{k} \vert = 0}  \rangle = \frac{1}{\sqrt{n_{A}}}\sum_{i_{A} \in A} \sum_{j \in \Lambda \setminus \{i_{A}\} } \vert j, i_{A} \rangle,
\end{equation}
where we use $\Lambda \setminus \{i_{A}\}$ to denote the set of all lattice sites excluding $i_{A}$.
Now observe that independent of $i_{A}$ exactly two of the terms in $\sum_{j \in \Lambda \setminus \{i_{A}\}} \vert j, i_{A} \rangle$ will match a term in the double summation in $\vert \mathcal{B}(1)_{ \vert \Vec{k} \vert = 0} \rangle = \frac{1}{\sqrt{3n_{B}}}\sum_{i_{B} \in B} \sum_{j_{A} \in n (i_{B})} \vert i_{B}, j_{A} \rangle$. We thus have 
\begin{equation}
    \langle \mathcal{B}(0)_{ \vert \Vec{k} \vert = 0} \vert h\sum_{i}\sigma^{x}_{i} \vert \mathcal{A}_{ \vert \Vec{k} \vert = 0}  \rangle = \frac{2n_{A}}{\sqrt{3n_{A}n_{B}}} = 2h\sqrt{\frac{n_{A}}{3n_{B}}}.
\end{equation}
We can make similar counting arguments (although we spare the full details) to give the full list of non-zero terms (not including their conjugates)
\begin{align}
    &\langle \mathcal{B}(0)_{ \vert \Vec{k} \vert = 0} \vert h\sum_{i}\sigma^{x}_{i} \vert \mathcal{A}_{ \vert \Vec{k} \vert = 0}  \rangle = 2h\sqrt{\frac{n_{A}}{3n_{B}}}, \notag \\
    &\langle \mathcal{B}(1)_{ \vert \Vec{k} \vert = 0} \vert h\sum_{i}\sigma^{x}_{i} \vert \mathcal{B}(0)_{ \vert \Vec{k} \vert = 0}  \rangle = \sqrt{3}h, \notag \\
    &\langle \mathcal{B}(2)_{ \vert \Vec{k} \vert = 0} \vert h\sum_{i}\sigma^{x}_{i} \vert \mathcal{B}(1)_{ \vert \Vec{k} \vert = 0}  \rangle = 2h, \notag \\
    &\langle \mathcal{B}(3)_{ \vert \Vec{k} \vert = 0} \vert h\sum_{i}\sigma^{x}_{i} \vert \mathcal{B}(2)_{ \vert \Vec{k} \vert = 0}  \rangle = \sqrt{3}h.
\end{align}
Lastly we consider the ratio $\sqrt{\frac{n_{A}}{n_{B}}}$. It is clear that for the finite unit cell (see Fig. \ref{Fig:iTNSHeavyHex}i) we have $\frac{n_{A}}{n_{B}} = \frac{3}{2}$. This ratio will hold true for any realisation of the periodic boundary heaxy hex lattice, as it is a simply a tiling of this unit cell with additional edges added to create the periodic boundary conditions. This gives us 
\begin{equation}
    \mathcal{P}_{\vert \Vec{k} \vert = 0} \left(h\sum_{i}\sigma^{x}_{i} \right) \mathcal{P}_{\vert \Vec{k} \vert = 0} = \begin{pmatrix}
        0 & 0 & \sqrt{2} h & 0 & 0 \\
        0 & 0 & \sqrt{3}h & 0 & 0 \\
        \sqrt{2}h & \sqrt{3}h & 0 &  0 & 2h \\
        0 & 0 & 0 & 0 & \sqrt{3}h \\
        0 & 0 & 2h & \sqrt{3}h & 0
    \end{pmatrix}
    \label{Eq:ReducedXHamiltonian}
\end{equation}
and finally 
\begin{equation}
    \mathcal{P}_{\vert \Vec{k} \vert = 0} H \mathcal{P}_{\vert \Vec{k} \vert = 0} = \mathcal{P}_{\vert \Vec{k} \vert = 0} \left(-J\sum_{\langle ij \rangle}\sigma^{z}_{i}\sigma^{z}_{j} + h\sum_{i}\sigma^{x}_{i} \right) \mathcal{P}_{\vert \Vec{k} \vert = 0} = \begin{pmatrix}
        4J & 0 & \sqrt{2} h & 0 & 0 \\
        0 & 6J & \sqrt{3}h & 0 & 0 \\
        \sqrt{2}h & \sqrt{3}h & 6J &  0 & 2h \\
        0 & 0 & 0 & 6J & \sqrt{3}h \\
        0 & 0 & 2h & \sqrt{3}h & 6J
    \end{pmatrix}
    \label{Eq:ReducedHamiltonianAppendix}
\end{equation}
as in the main text --- where we have ignored the constant $E_{0}$ on the diagonal as it simply shifts all the eigenvalues and is irrelevant.

\bibliography{Bibliography}

\end{document}